\documentclass[letter]{aa} 
\usepackage{graphicx}
\usepackage{txfonts}
\usepackage{natbib}
\usepackage{longtable}

\bibpunct[]{(}{)}{;}{a}{}{,}
\begin{document}
   \title{Spectroscopic parameters and rest frequencies of\\ isotopic methylidynium, CH$^+$}

   \author{Holger~S.~P. M{\"u}ller\inst{}}

   \institute{I.~Physikalisches Institut, Universit{\"a}t zu K{\"o}ln,
              Z{\"u}lpicher Str. 77, 50937 K{\"o}ln, Germany\\
              \email{hspm@ph1.uni-koeln.de}}

   \date{Received 10 March 2010 / Accepted 26 April 2010}

  \abstract
{Astronomical observations toward Sagittarius~B2(M) as well as other sources 
with APEX have recently suggested that the rest frequency of the $J = 1 - 0$ 
transitions of $^{13}$CH$^+$ is too low by about 80~MHz.}
{Improved rest frequencies of isotopologs of methylidynium should be derived 
to support analyses of spectral recording obtained with the ongoing $Herschel$ mission 
or the upcoming SOFIA.}
{Laboratory electronic spectra of four isotopologs of CH$^+$ have been subjected 
to one global least-squares fit.
Laboratory data for the $J =  1 - 0$ ground state rotational transitions of CH$^+$, 
$^{13}$CH$^+$, and CD$^+$, which became available during the refereeing process, 
have been included in the fit as well.}
{An accurate set of spectroscopic parameters has been obtained together with 
equilibrium bond lengths and accurate rest frequencies for six CH$^+$ isotopologs: 
CH$^+$, $^{13}$CH$^+$, $^{13}$CD$^+$, CD$^+$, $^{14}$CH$^+$, and CT$^+$.}
{The present data will be useful for the analyses of $Herschel$ or SOFIA observations 
of methylidynium isotopic species.}

\titlerunning{rest frequencies of isotopic CH$^+$}

\maketitle
\hyphenation{Sym-po-si-um}
\hyphenation{Co-lum-bus}

%

\section{Introduction}
\label{intro}

The first three molecules detected in space in the 1930s were 
methylidynium, CH$^+$, methylidyne, CH, and cyanogen, CN. 
\citet{CH+_ident} identified three absorption lines, which had been observed 
toward several stars, as belonging to the $A\ ^1\Pi - X\ ^1\Sigma ^+$ 
electronic transition of CH$^+$. They were assigned to the $R(0)$ 
($\equiv J = 1 - 0$) transitions of the $\varv = 0 - 0$, $1 - 0$, 
and $2 - 0$ vibrational bands. 

The $J = 2 - 1$, $3 - 2$, and $4 - 3$ rotational transitions were identified 
by \citet{CH+_NGC7027} in spectra recorded with the Long Wavelength Spectrometer 
of the Infrared Space Observatory toward the planetary nebula NGC~7027. 
The observation of the $J = 1 - 0$ rotational transitions at 835137.5~MHz 
\citep{isos-CH+_rot} is hampered by a telluric line of molecular oxygen close by 
\citep{O2_rot_2010} and thus requires observations from space, e.g. with 
the recently launched {\it Herschel} satellite or with the Stratospheric 
Observatory For Infrared Astronomy (SOFIA).

Recently, \citet{det_13CH+} reported on the observation of the 
$J = 1 - 0$ transition of $^{13}$CH$^+$ in absorption toward the star-forming 
region G10.6-0.4. They deduced 830132~(3)~MHz as the prefered rest frequency 
based on scaling from the CH$^+$ laboratory rest frequency. Alternative 
rest frequencies of 830107~(1) or 830193~(4)~MHz were excluded. 
Very recently, \citet{hydrides_2010} carried out an absorption study 
with the Atacama Pathfinder EXperiment (APEX) 
of light hydride species toward the evolved massive star-forming region 
Sagittarius (Sgr for short) B2(M). They detected low-lying rotational 
transitions of $^{13}$CH$^+$, H$^{35}$Cl, H$^{37}$Cl, and, for the first time, 
of SH$^+$. Similar studies have been carried out toward additional 
sources (F. Wyrowski, private communication). 
A comparison of the line profiles among these species together 
with observations of additional species at lower frequencies suggested 
a rest frequency of about 830210~MHz, approximately 80~MHz higher 
than the prefered rest frequency from \citet{det_13CH+}, but in reasonable 
agreement with an alternative rest frequency of 830193~(4)~MHz 
they had dismissed. 

\citet{CH+_rot} obtained a value of 835078.950~(75)~MHz as the $J = 1 - 0$ 
transition frequency of CH$^+$, about 5~MHz lower than they deduced from 
analyses of the $A\ ^1\Pi - X\ ^1\Sigma ^+$ electronic spectra of four 
isotopologs \citep{12CH+_1982,13CH+_UV_1997,CD+_UV_1987,13CD+_UV_1997}. 

During the refereeing process of this manuscript I received 
$J = 1 - 0$ transition frequencies for CH$^+$, $^{13}$CH$^+$, and CD$^+$, 
very recently obtained by \citet{isos-CH+_rot}. 
While the latter two frequencies ($\sim$830216 and 453522~MHz) 
were rather close to predictions made from the spectroscopic parameters 
of the respective isotopolog \citep{13CH+_UV_1997,CD+_UV_1987}, 
the CH$^+$ frequency (835137.5~MHz) was almost 40~MHz higher 
than that deduced from a more recent analysis of the CH$^+$ 
electronic spectrum \citep{CH+_UV_2006}, but almost 70~MHz higher 
than measurements by \citet{CH+_rot} indicated. 
Because no further rotational data and no infrared transitions 
have been reported thus far, determinations of methylidynium rest 
frequencies still rely quite heavily on the electronic spectra.

In the current work, I present a combined reanalysis of the 
$A\ ^1\Pi - X\ ^1\Sigma ^+$ electronic spectra of four isotopic species 
of CH$^+$ together with the available rotational transitions 
to derive reliable rest frequencies for CH$^+$ isotopologs.

\section{Analysis and discussion}
\label{a_and_d}

The ro-vibrational energy levels of a diatomic molecule AB 
can be represented by the Dunham expression

\begin{equation}
\label{Dunham}
E(\varv, J)/h = \sum_{i,j} Y_{ij}(\varv + 1/2)^i J^j (J + 1)^j,
\end{equation}

where the $Y_{ij}$ are the Dunham parameters. In electronic states 
different from $\Sigma$ states, i.e. in states with orbital angular momentum 
$\Lambda > 0$, the expansion in $J(J + 1)$ is commonly replaced by an 
expansion in $J(J + 1) - \Lambda ^2$, see e.g. \citep{Dunham_BO_Watson2}. 
Obviously, the two are the same for $\Sigma$ states with $\Lambda = 0$, 
such as the ground electronic state of CH$^+$. 
The $A$ state is a $^1\Pi$ electronic state with $\Lambda = 1$, 
and the expansion has been carried out in $J(J + 1) - 1$; 
an expansion in $J(J + 1)$ can be found also, see e.g. the case of the 
$^2 \Pi$ radical BrO \citep{BrO_rot}.

The energy levels in a $^1 \Pi$ electronic state are modified by 
a $\lambda$-doubling which can be expressed as

\begin{equation}
\label{lambda}
E^\lambda (\varv, J)/h = \pm J(J+1)/2 \sum_{i,j} q_{ij}(\varv + 1/2)^i J^j (J + 1)^j,
\end{equation}

where the $\lambda$-doubling parameters $q_{ij}$ scale with $\mu _c^{-2-(i+2j)/2}$
\citep{radi-Hamiltonian}. 

\citet{Dunham_BO_Watson1,Dunham_BO_Watson2} has shown that several 
isotopic species of AB can be fit jointly by constraining the $Y_{ij}$ to

\begin{equation}
\label{BO}
Y_{i,j} = U_{i,j} \left( 1 + \frac{m_e {\it \Delta}^{\rm A}_{ij}}{M_{\rm A}} 
 + \frac{m_e {\it \Delta}^{\rm B}_{ij}}{M_{\rm B}}\right) \mu _c^{-(i+2j)/2}, 
\end{equation}

where $U_{i,j}$ is isotope invariant, $m_e$ is the mass of the electron, 
$\mu_c = M_{\rm A}M_{\rm B}/(M_{\rm A} + M_{\rm B} - nm_e)$ is the 
charge-corrected reduced mass of AB and $n$ is the charge of AB, 
$M_{\rm A}$ is the mass of atom A, and ${\it \Delta}^{\rm A}_{ij}$ 
is a Born-Oppenheimer breakdown (BOB) term.
$U_{ij} \mu^{-(i+2j)/2} {\it \Delta}_{ij}^{\rm A} m_e/M_{\rm A}$ is sometimes 
abbreviated as $\delta_{ij}^{\rm A}$. Moreover, it is noteworthy 
that both ${\it \Delta}_{ij}^{\rm A}$ and $\delta_{ij}^{\rm A}$ are 
defined negatively in some papers. Obviously, ${\it \Delta}_{ij}^{\rm B}$ 
and $\delta_{ij}^{\rm B}$ are defined equivalently.

A specific transition of the  $A\ ^1\Pi - X\ ^1\Sigma ^+$ electronic spectrum 
of CH$^+$ is well described by the isotopolog, the vibrational quantum 
numbers in the upper and lower electronic states, the lower state rotational 
quantum number $J$ and whether the transition belongs to the $R$, $Q$, or 
$P$-branch in which the upper state rotational quantum number is $J + 1$, $J$, 
and $J -  1$, respectively; $J$ is commonly given in parentheses.


\begin{table}
\begin{center}
\caption{Experimentally determined spectroscopic parameters$^a$ 
(cm$^{-1}$, MHz) of methylidynium, $^{12}$CH$^+$ in its 
$X\ ^1\Sigma ^+$ and $A\ ^1\Pi$ electronic states from a global fit 
involving four isotopologs.}
\label{det-parameter}
{
\renewcommand{\arraystretch}{1.07}
\begin{tabular}[t]{lr@{}l}
\hline \hline
Parameter & \multicolumn{2}{c}{Value} \\
\hline
\multicolumn{3}{c}{$X\ ^1\Sigma ^+$} \\
\hline
$U_{10} \mu ^{-1/2}$                                              &     2\,860&.387~(22)      \\
$U_{10} \mu ^{-1/2}{\it \Delta} ^{\rm C} _{10} m_e/M_{\rm C}$     &       $-$0&.147~(18)      \\
$U_{10} \mu ^{-1/2}{\it \Delta} ^{\rm H} _{10} m_e/M_{\rm H}$     &       $-$2&.672~(13)      \\
$U_{20} \mu ^{-1}$                                                &      $-$59&.369\,7~(61)   \\
$U_{20} \mu ^{-1}{\it \Delta} ^{\rm H} _{20} m_e/M_{\rm H}$       &          0&.051\,0~(62)   \\
$Y_{30}$                                                          &          0&.225\,46~(106) \\
$U_{01} \mu ^{-1}$                                                &   427\,360&.68~(121)      \\
$U_{01} \mu ^{-1}{\it \Delta} ^{\rm C} _{01} m_e/M_{\rm C}$       &     $-$155&.80~(21)       \\
$U_{01} \mu ^{-1}{\it \Delta} ^{\rm H} _{01} m_e/M_{\rm H}$       &  $-$2\,146&.22~(177)      \\
$U_{11} \mu ^{-3/2}$                                              & $-$14\,917&.7~(57)        \\
$U_{11} \mu ^{-3/2}{\it \Delta} ^{\rm H} _{11} m_e/M_{\rm H}$     &         58&.6~(27)        \\
$Y_{21}$                                                          &         91&.8~(41)        \\
$Y_{31}$                                                          &       $-$3&.84~(66)       \\
$U_{02} \mu ^{-2}$                                                &      $-$42&.468~(21)      \\
$U_{02} \mu ^{-2}{\it \Delta} ^{\rm H} _{02} m_e/M_{\rm H}$       &          0&.563~(20)      \\
$Y_{12}$                                                          &          0&.960~(34)      \\
$Y_{22}$                                                          &       $-$0&.045~(10)      \\
$Y_{03} \times 10^3$                                              &          4&.264~(86)      \\
$Y_{13} \times 10^3$                                              &       $-$0&.726~(53)      \\
$Y_{04} \times 10^6$$^c$                                          &       $-$0&.4             \\
$C(^{13}$C)                                                       &          1&.091~(28)      \\
\hline
\multicolumn{3}{c}{$A\ ^1\Pi$} \\
\hline
$U_{00}$(H)$^b$                                                    &    23\,639&.326~(23)      \\
$U_{00}{\it \Delta} ^{\rm C} _{10} m_e/M_{\rm C}$(H)$^b$           &      $-$19&.620~(24)      \\
$U_{00}$(D)$^b$                                                    &    23\,786&.537~(24)      \\
$U_{00}{\it \Delta} ^{\rm C} _{10} m_e/M_{\rm C}$(D)$^b$           &      $-$26&.436~(26)      \\
$U_{10} \mu _c^{-1/2}$(H)                                          &     1\,864&.379~(14)      \\
$U_{10} \mu _c^{-1/2}{\it \Delta} ^{\rm C} _{10} m_e/M_{\rm C}$(H) &          0&.106~(12)      \\
$U_{10} \mu _c^{-1/2}$(D)                                          &     1\,367&.473~(17)      \\
$U_{10} \mu _c^{-1/2}{\it \Delta} ^{\rm C} _{10} m_e/M_{\rm C}$(D) &       $-$0&.180~(19)      \\
$Y_{20}$(H)                                                        &     $-$115&.884\,7~(48)   \\
$Y_{20}$(D)                                                        &      $-$60&.815\,4~(47)   \\
$Y_{30}$(H)                                                        &          2&.639\,48~(85)  \\
$Y_{30}$(D)                                                        &          0&.776\,92~(78)  \\
$U_{01} \mu _c^{-1}$                                               &   356\,717&.5~(63)        \\
$U_{01} \mu _c^{-1}{\it \Delta} ^{\rm C} _{01} m_e/M_{\rm C}$      &       $-$9&.2~(53)        \\
$U_{01} \mu _c^{-1}{\it \Delta} ^{\rm H} _{01} m_e/M_{\rm H}$      &     $-$316&.4~(24)        \\
$U_{11} \mu _c^{-3/2}$                                             & $-$27\,562&.7~(70)        \\
$U_{11} \mu _c^{-3/2}{\it \Delta} ^{\rm H} _{11} m_e/M_{\rm H}$    &         24&.06~(147)      \\
$Y_{21}$                                                           &     $-$634&.8~(49)        \\
$Y_{31}$                                                           &        137&.76~(95)       \\
$Y_{02}$                                                           &      $-$58&.166~(26)      \\
$Y_{12}$                                                           &       $-$2&.502~(61)      \\
$Y_{22}$                                                           &       $-$0&.252~(44)      \\
$Y_{32}$                                                           &          0&.149\,3~(87)   \\
$Y_{03} \times 10^3$                                               &       $-$0&.876~(84)      \\
$Y_{13} \times 10^3$                                               &       $-$0&.911~(63)      \\
$q_{00}$                                                           &     1\,234&.25~(108)      \\
$q_{10}$                                                           &      $-$96&.53~(82)       \\
$q_{01}$                                                           &       $-$0&.643\,6~(98)   \\
$q_{11}$                                                           &       $-$0&.020\,3~(79)   \\
\hline
\end{tabular}
}
\end{center}
$^a$ The vibrational parameters $Y_{i0}$ as well as related parameters 
are given in cm$^{-1}$; all other $Y_{ij}$ as well as related parameters 
are given in MHz. Numbers in parentheses are one standard deviation in 
units of the least significant figures.\\
$^b$ Determined with respect to the lowest allowed energy levels of 
each electronic state.\\
$^c$ Estimated, see text.
\end{table}


None of the investigations of the $A - X$ electronic spectrum of CH$^+$ 
and its isotopologs provided uncertainties for individual lines, 
only estimates of average precision and accuracy. 
In the present work the initial uncertainties were estimated for each 
isotopolog and each vibronic band separately based on average 
residuals between observed frequencies and those calculated 
from the spectroscopic parameters. 
During the analysis, the uncertainties were adjusted 
to yield an rms error very close to the theoretical value of 1.0, 
usually slightly lower. Transitions with residuals between observed 
and calculated frequencies much higher than the average were omitted 
from the fit. These may have been weak lines with very low or very high 
values of $J$ or accidentally overlapped lines.
For each isotopolog and each vibronic band the finally ascribed uncertainty 
will be given below in parentheses in units of 0.001~cm$^{-1}$ 
together with the lines from the original literature, which 
have been omitted from the fit. Lines already discarded in the original analyses 
were usually omitted here as well; these lines will not be mentioned specifically.

The fit was started with the CH$^+$ data from \citet{12CH+_1982} and 
\citet{CH+_UV_2006}, and the initial parameters were taken from the latter work. 
As the $Y_{0i}$ got smaller by factors of $\sim -10000$, $Y_{04}$ was 
estimated as $-0.4$~Hz.
\citet{CH+_UV_2006} measured the $\varv = 0 - 0$ (5 ($\times0.001$~cm$^{-1}$), 
$R$(17)), $0 - 1$ (5, $-$), and $2 - 1$ (9, $P$(14)) bands, 
while \citet{12CH+_1982} obtained data for the $\varv = 0 - 0$ (5, $P$(6)), 
$0 - 1$ (6, $Q$(14)), $1 - 0$ (12, $R$(3)), $1 - 1$ (7, $P$(7,8)), 
$1 - 2$ (6, $P$(7)), $1 - 3$ (4, $R$(5)), $2 - 1$ (5, $-$), and 
$3 - 1$ (7, $Q$(9), $P$(2)) bands.
The CH$^+$ $J = 1 - 0$ rotational transition predicted from this data 
set is $835093.42 \pm 5.25$~MHz, almost exactly 10~MHz higher than what 
\citet{CH+_rot} obtained and about 14.5~MHz higher than the line these authors 
measured in the laboratory. 
Inclusion of data from three additional isotopologs, as described below, 
shifted the prediction to somewhat higher frequencies, $\sim$835112~MHz 
being the highest as long the the weights of the CH$^+$ data were not reduced.

Scaling these parameters directly with $\mu _c$ yields 830146.97~MHz for 
the $^{13}$CH$^+$, $J = 1 - 0$ transition. In contrast, predictions from 
\citet{13CH+_UV_1997} yield a value approximately 80~MHz higher, suggesting 
a large BOB term for $Y_{01}$ will be necessary at least in the ground 
electronic state. Inclusion of data from the $\varv = 0 - 0$ (4, $-$) band 
confirmed this.
These data also required a BOB term for $Y_{01}$ as well as for the term value 
$T$ of the excited electronic $A$ state, though the latter value appeared to 
be too high to be considered a BOB term in the usual sense. 
The $J = 1 - 0$ ground state rotational transition was now predicted at 
$830206.82 \pm 5.85$~MHz, 60 MHz higher than from direct scaling as a result 
of large effects caused by the breakdown of the Born-Oppenheimer 
approximation. The $\varv = 0 - 1$ (4, $-$), $2 - 0$ (8, $R$(8)) 
and $2 - 1$ (12, $-$) bands were included without much difficulty. 
In the $\varv = 1 - 0$ (6, $-$) band, in which already several lines 
were not used in the original fits, the $R$(5,9) and $P$(3,4,10) lines 
were omitted in addition in the present work. 
In the $\varv = 1 - 1$ (9, $-$) band, $R$(6,7) are in fact $R$(7,8), 
similarly the $J$ values have to be revised one up for $Q$(7,8). 
Moreover, $R$(7,9,10) were omitted from the fit. 
With more vibronic bands included, a correction term for $Y_{10}$ 
was also employed in the fit. 

The CD$^+$ data from \citet{CD+_UV_1987} proved to be much more difficult 
to reproduce. Not only were BOB terms required for the $Y_{01}$ and 
$Y_{10}$, as for $^{13}$CH$^+$, but also for $Y_{20}$, $Y_{11}$, and 
$Y_{02}$ in both electronic states. The correction term to $T$ turned out 
to be very large again. The $\varv = 0 - 0$ (4, $-$), $0 - 1$ (4, $R$(12)), 
$1 - 0$ (3, $R$(11)), $P$(2,6,12,13)), $1 - 2$ (9, $-$), $2 - 1$ (6, $-$), 
and $3 - 1$ (5, $Q$(1), $P$(6,11)) bands were used in the fit.  
The $\varv = 1 - 3$ band was measured but not used in the fit by 
\citet{CD+_UV_1987}, it is on average about 0.09~cm$^{-1}$ too low. 
In addition, $\varv = 2 - 0$ was also not used in the present fit as 
it was about 0.03~cm$^{-1}$ too low. Its inclusion in the fit would have 
afforded {\it both} $\varv = 2 - 1$ and $\varv = 3 - 1$ to be omitted 
from the fit.


\begin{table}
\begin{center}
\caption{Derived spectroscopic parameters$^a$ (cm$^{-1}$, MHz, unitless) 
of methylidynium, $^{12}$CH$^+$ in its $X\ ^1\Sigma ^+$ and $A\ ^1\Pi$ 
electronic states from a global fit involving four isotopologs.}
\label{der-parameter}
{
\renewcommand{\arraystretch}{1.10}
\begin{tabular}[t]{lr@{}l}
\hline \hline
Parameter & \multicolumn{2}{c}{Value} \\
\hline
\multicolumn{3}{c}{$X\ ^1\Sigma ^+$} \\
\hline
$Y_{10}$                                                      &     2\,857&.569\,2~(92)   \\
${\it \Delta} ^{\rm C} _{10}$                                 &       $-$1&.120~(137)     \\
${\it \Delta} ^{\rm H} _{10}$                                 &       $-$1&.715\,9~(86)   \\
$Y_{20}$                                                      &      $-$59&.318\,7~(60)   \\
${\it \Delta} ^{\rm H} _{20}$                                 &       $-$1&.578~(188)     \\
$Y_{01}$                                                      &   425\,058&.66~(248)      \\
${\it \Delta} ^{\rm C} _{01}$                                 &       $-$7&.974\,9~(105)  \\
${\it \Delta} ^{\rm H} _{01}$                                 &       $-$9&.226\,3~(76)   \\
$Y_{11}$                                                      & $-$14\,859&.0~(67)        \\
${\it \Delta} ^{\rm H} _{11}$                                 &       $-$7&.222~(33)      \\
$Y_{02}$                                                      &      $-$41&.905~(20)      \\
${\it \Delta} ^{\rm H} _{02}$                                 &      $-$24&.36~(86)       \\
\hline
\multicolumn{3}{c}{$A\ ^1\Pi$} \\
\hline
$T$(H)$^b$                                                    &    23\,619&.707~(5)       \\
$T$(D)$^b$                                                    &    23\,760&.102~(5)       \\
$Y_{10}$(H)                                                   &     1\,864&.485~(8)       \\
${\it \Delta} ^{\rm C} _{10}$(H)                              &          1&.244~(145)     \\
$Y_{10}$(D)                                                   &     1\,367&.294~(8)       \\
${\it \Delta} ^{\rm C} _{10}$(D)                              &       $-$2&.876~(306)     \\
$Y_{01}$                                                      &   356\,391&.8~(29)        \\
${\it \Delta} ^{\rm C} _{01}$                                 &       $-$0&.56~(33)       \\
${\it \Delta} ^{\rm H} _{01}$                                 &       $-$1&.630~(12)      \\
$Y_{11}$                                                      & $-$27\,538&.7~(71)        \\
${\it \Delta} ^{\rm H} _{11}$                                 &       $-$1&.604~(98)      \\
\hline
\end{tabular}
}
\end{center}
$^a$ The electronic term values $T$, and the vibrational parameters 
$Y_{i0}$ are given in cm$^{-1}$; all other $Y_{ij}$ are given in MHz; 
the ${\it \Delta} ^{\rm X} _{ij}$ are unitless. Numbers in parentheses 
are one standard deviation in units of the least significant figures.\\
$^b$ Determined with respect to the lowest allowed energy levels of 
each electronic state.
\end{table}


Lastly, $^{13}$CD$^+$ data from \citet{13CD+_UV_1997} were included 
in the fit, namely the $\varv = 0 - 0$ (6, $P$(12,13,14)) and $1 - 0$ 
(4, $-$) bands. The band origins were not quite compatible with the other 
data. In combination with the considerable corrections required for $T$, 
this was interpreted as a sign of the perturbation of the $A$ electronic state. 
Therefore, separate term values as well as $Y_{i0}$ were used for CH$^+$ and 
CD$^+$. In addition, different carbon BOB terms were required for $T$ 
and $Y_{10}$.

The referee of this manuscript pointed out new laboratory measurements 
of the $J = 1 - 0$ rotational transitions of CH$^+$, $^{13}$CH$^+$, and 
CD$^+$ which I received soon thereafter from the author \citep{isos-CH+_rot}. 
The $^{13}$CH$^+$ and CD$^+$ frequencies were close to the predictions based 
on parameters from \citet{13CH+_UV_1997,CD+_UV_1987} as well as predictions 
from present fits which only included data from the electronic spectra. 
In contrast, the CH$^+$ frequency was higher than predictions from the 
current fits by between 25 and 45~MHz. 
The CH$^+$ and $^{13}$CH$^+$ lines showed large Zeeman 
splitting in a magnetic field, which, as already mentioned by 
\citet{isos-CH+_rot}, requires a high $g_J$-value for 
a $^1 \Sigma ^+$ molecule. This in turn leads to high $\Delta _{01}$ 
BOB terms for CH$^+$ in the ground electronic state, see below. 
The large $g_J$-value is also compatible with the large $^{13}$C nuclear 
spin-rotation splitting of 1.6~MHz in the $J = 1 - 0$ transition 
of $^{13}$CH$^+$, as already pointed out by \citet{isos-CH+_rot}.

The fit easily accomodated these rotational data; the rms error of the fit 
only slightly increased from 0.944 to 0.991. 
This rather small deterioration of the fit may indicate unfavorable 
correlation of the parameters, which is reduced by the inclusion of the 
pure rotational data in the fit. 
Weighting two vibronic bands of CH$^+$ each from \citet{12CH+_1982,CH+_UV_2006} 
slightly lower ($\leq 25$\,\%) yielded rms errors of all vibronic bands 
very close to 1.0 and an overall rms error of 0.945. 
The inclusion of the rotational data caused the $^{13}$C BOB correction 
value to $U_{01} \mu ^{-1}$ in the ground electronic state to be reduced 
by almost half, similar changes in magnitude occured in the excited $A$ 
state $^{13}$C BOB term. These changes were mostly compensated by changes 
in $U_{01} \mu ^{-1}$, to a much lesser extent by the BOB term for H. 
Changes considerably outside the uncertainties also occured for 
$U_{02} \mu ^{-2}$ and $Y_{02}$ in the $X$ and $A$ state, respectively.
All other changes were mostly well within the uncertainties.

The final spectroscopic parameters are given in Table~\ref{det-parameter}; 
further spectroscopic parameters, which were derived from these parameters, 
are given in Table~\ref{der-parameter}. Predictions of the $J = 1 - 0$ 
and $2 - 1$ transitions of six isotopologs are given in Table~\ref{pred:rot}  
together with the measured ones; 
these should be useful for astronomical observations 
as well as for laboratory spectroscopic investigations. 
The three laboratory rotational transitions \citep{isos-CH+_rot} determine 
three parameters related to $U_{01}$. In addition, the weight of each line 
in the fit is much larger than that of a rovibronic transition. Therefore, 
it is not surprising that in these cases predicted frequencies and 
their uncertainties are essentially identical to laboratory values 
\citep{isos-CH+_rot}. 
More extensive predictions, including those of rovibrational spectra, 
as well as details of the fit file will be available in the 
Cologne Database for Molecular Spectroscopy, 
CDMS,\footnote{website: http://www.astro.uni-koeln.de/cdms/} 
\citep{CDMS_1,CDMS_2}. 
Extrapolations beyond $J = 1 - 0$ should be viewed with some caution, 
meaning the actually measured transition frequency may be more than three 
times the predicted uncertainty away from the calculated position, 
especially for the CH$^+$ main isotopolog because the rotational transition 
was rather far away from prediction absed on electronic spectra.
Predictions with $J \ge 10$ should be viewed with great caution.


\begin{table}
\begin{center}
\caption{Measured \citep{isos-CH+_rot} and calculated low-$J$ rotational 
transition frequencies$^a$ (MHz) of six isotopic species of CH$^+$.}
\label{pred:rot}
\begin{tabular}[t]{lr@{}lr@{}lr@{}l}
\hline \hline
 & \multicolumn{2}{c}{Measured} & \multicolumn{4}{c}{Calculated} \\
\cline{4-7}
 & \multicolumn{2}{c}{$J = 1 - 0$} & \multicolumn{2}{c}{$J = 1 - 0$} & \multicolumn{2}{c}{$J = 2 - 1$} \\
\hline
CH$^+$        & 835\,137&.504~(20)     & 835\,137&.503~(20) & 1\,669\,281&.3~(3) \\
$^{13}$CH$^+$ & 830\,216&.095~(30)$^b$ & 830\,216&.096~(22) & 1\,659\,450&.3~(3) \\
CD$^+$        & 453\,521&.851~(20)     & 453\,521&.851~(20) &    906\,752&.0~(1) \\
$^{13}$CD$^+$ &         &              & 448\,539&.258~(26) &    896\,793&.2~(1) \\
$^{14}$CH$^+$ &         &              & 826\,012&.748~(45) & 1\,651\,053&.6~(3) \\
CT$^+$        &         &              & 325\,443&.364~(70) &    650\,736&.8~(1) \\
\hline
\end{tabular}
\end{center}
$^a$ Numbers in parentheses are one standard deviation in units of the 
least significant figures; see the text for discussion of uncertainties.\\
$^b$ Value corrected for hyperfine splitting; two lines were measured, see text, 
each accurate to 30~kHz.
\end{table}


In contrast to e.g. \citet{12CH+_1982}, but in agreement with 
\citet{CH+_UV_2006}, the excited $A$ electronic state could be reproduced 
in a rather satisfactorily manner using power series in $\varv + 1/2$ 
besides $J(J + 1) - 1$ for each isotopolog individually and, 
with the exception of the term value and the vibrational spacings, 
also for the combined data set of four isotopic species. 
The fact that separate parameters had to be used for CH$^+$ and 
CD$^+$ may be explained by a perturbation of the $A$ state. 
Essentially all parameters $Y_{ij}$ with $j > 0$ or related parameters 
$U_{i,j} \mu _c^{-(i+2j)/2}$ in Table~\ref{det-parameter} 
as well as the BOB terms in Table~\ref{der-parameter}
appear to be reasonable. 

The $Y_{ij}$ or related parameters for the $X$ ground electronic state 
in Table~\ref{det-parameter} again appear rather reasonable. 
The BOB terms ${\it \Delta} _{i0}$ in Table~\ref{der-parameter} are 
negative and their magnitudes are normal 
\citep{Dunham_BO_Watson1,Dunham_BO_Watson2,BrO_rot,CDMS_2,SiS_rot}. 
The BOB terms ${\it \Delta} _{i1}$, however, are rather large,
but of similar magnitude. 
\citet{Dunham_BO_Watson1} has shown that ${\it \Delta} _{01}$ has one 
contribution $g_J m_e/m_p$, where $g_J$ is the rotational $g$-factor 
and $m_p$ is the mass of the proton, which may get big for 
low-lying electronic states of the same multiplicity, i.e. 
singlet states in the case of CH$^+$. Because $Y_{11}$ is the first 
vibrational correction term to $Y_{01}$ it is probably reasonable that 
${\it \Delta} ^{\rm H} _{11}$ is large in magnitude also. 
\citet{BO-initio} have shown that ${\it \Delta} _{01}$ can be calculated 
ab initio and presented results which were compared with experimental 
data for SiS \citep{SiS_rot}, HCl, and HF. 
Calculations for CH$^+$ will be very welcome.

Equilibrium structural parameters of 112.77543~(16) and 123.45370~(106)~pm 
have been derived for the $X$ and $A$ electronic state, respectively, 
in the BO approximation which are very different from 
113.08026~(33) and 123.50991~(50)~pm obtained for the CH$^+$ isotopolog 
under consideration of the breakdown of the BO approximation. 
The latter values agree fairly well with 113.08843~(30) and 
123.5053~(37)~pm obtained by \citet{CH+_UV_2006}. 

\section{Conclusion}
\label{conclusion}

The $A\ ^1\Pi - X\ ^1\Sigma ^+$ electronic spectra of four isotopologs 
plus the rotational data for three species have been reproduced 
rather well with one set of spectroscopic parameters and 
with only comparatively few transitions omitted from the fit. 
Considering that the $A$ electronic state has been judged to be 
perturbed by an unidentified state, the spectroscopic parameters show 
few peculiarities. 

The predicted $J = 1 - 0$ transition frequencies of the three remaining 
isotopologs as well as higher-$J$ predictions for all species will be 
useful for future astronomical observations or laboratory spectroscopic 
investigations. The data for $^{14}$CH$^+$ and CT$^+$ may be useful 
for determining if $^{14}$C or T play a significant role in certain 
(circumstellar) environments. 


\begin{acknowledgements}
I thank Karl M. Menten for comments on the manuscript, 
the referee for pointing out very new laboratory measurements on  
CH$^+$ isotopologs, and Takayoshi Amano for providing these 
prior to submission. 
I am also very grateful to the Bundesministerium f\"ur Bildung und 
Forschung (BMBF) for financial support aimed at maintaining the 
Cologne Database for Molecular Spectroscopy, CDMS. This support has been 
administered by the Deutsches Zentrum f\"ur Luft- und Raumfahrt (DLR). 
\end{acknowledgements}




\begin{thebibliography}{}

\bibitem[Amano(2010)]{isos-CH+_rot} 
Amano, T. 
2010, \apjl, accepted

\bibitem[Bembenek et al.(1987)]{CD+_UV_1987} 
Bembenek, Z., Cisak, H., \& K{\c e}pa, R. 
1987, J. Phys. B, 20, 6197 

\bibitem[Bembenek(1997a)]{13CH+_UV_1997} 
Bembenek, Z. 
1997, J. Mol. Spectrosc., 181, 136 

\bibitem[Bembenek(1997b)]{13CD+_UV_1997} 
Bembenek, Z. 
1997, J. Mol. Spectrosc., 182, 439 

\bibitem[Brown et al.(1979)]{radi-Hamiltonian} 
Brown, J.~M., Colbourn, E.~A., Watson, J.~K.~G., \& Wayne, F.~D. 
1979, J. Mol. Spectrosc., 74, 294 

\bibitem[Carrington \& Ramsay(1982)]{12CH+_1982} 
Carrington, A., \& Ramsay, D.~A. 
1982, \physscr, 25, 272 

\bibitem[Cernicharo et al.(1997)]{CH+_NGC7027} 
Cernicharo, J., Liu, X.-W., Gonzalez-Alfonso, E., et al.
1997, \apjl, 483, L65 

\bibitem[Douglas \& Herzberg(1941)]{CH+_ident} 
Douglas, A.~E., \& Herzberg, G. 
1941, \apj, 94, 381

\bibitem[Drouin et al.(2001)]{BrO_rot} 
Drouin, B.~J., Miller, C.~E., M{\"u}ller, H.~S.~P., \& Cohen, E.~A. 
2001, J. Mol. Spectrosc., 205, 128 

\bibitem[Drouin et al.(2010)]{O2_rot_2010}
Drouin, B.~J., Yu, S., Miller, C.~E., et al.
2010 \jqsrt, 111, 1167; 
and references therein

\bibitem[Falgarone et al.(2005)]{det_13CH+} 
Falgarone, E., Phillips, T.~G., \& Pearson, J.~C. 
2005, \apjl, 634, L149 

\bibitem[Gauss \& Puzzarini(2010)]{BO-initio}
Gauss, J., \& Puzzarini, C. 
2010, Mol. Phys., 108, 269 

\bibitem[Hakalla et al.(2006)]{CH+_UV_2006} 
Hakalla, R., K{\c e}pa, R., Szajna, W., \& Zachwieja, M. 
2006, Eur. Phys. J. D, 38, 481 

\bibitem[Menten et al.(2010)]{hydrides_2010}
Menten, K.~M., Wyrowski, F., Belloche, A., et al.
2010, \aap, submitted

\bibitem[M{\"u}ller et al.(2001)]{CDMS_1}
M{\"u}ller, H.~S.~P., Thorwirth, S., Roth, D.~A.,
\& Winnewisser, G.
2001, A\&A, 370, L49

\bibitem[M{\"u}ller et al.(2005)]{CDMS_2}
M{\"u}ller, H.~S.~P., Schl{\"o}der, F., Stutzki, J.,
\& Winnewisser, G.
2005, J. Mol. Struct, 742, 215

\bibitem[M{\"u}ller et al.(2007)]{SiS_rot} 
M{\"u}ller, H.~S.~P., McCarthy, M.~C., Bizzocchi, L., et al. 
2007, Phys. Chem. Chem. Phys., 9, 1579 

\bibitem[Pearson \& Drouin(2006)]{CH+_rot} 
Pearson, J.~C., \& Drouin, B.~J. 
2006, \apjl, 647, L83 

\bibitem[Watson(1973)]{Dunham_BO_Watson1} 
Watson, J.~K.~G. 
1973, J. Mol. Spectrosc., 45, 99 

\bibitem[Watson(1980)]{Dunham_BO_Watson2} 
Watson, J.~K.~G. 
1980, J. Mol. Spectrosc., 80, 411 


\end{thebibliography}
\end{document}